\documentclass[pdflatex,sn-apa]{sn-jnl}

\usepackage[table]{xcolor}
\definecolor{aquablue}{cmyk}{0.12,0,0.04,0}

\usepackage{graphicx}%
\usepackage{multirow}%
\usepackage{amsmath,amssymb,amsfonts}%
\usepackage{amsthm}%
\usepackage{mathrsfs}%
\usepackage[title]{appendix}%
\usepackage{xcolor}%
\usepackage{textcomp}%
\usepackage{manyfoot}%
\usepackage{booktabs}%
\usepackage{algorithm}%
\usepackage{algorithmicx}%
\usepackage{algpseudocode}%
\usepackage{listings}%

\let\cline\cmidrule

\usepackage{placeins}
\usepackage{soul}
\colorlet{usercolorname}{yellow!0}
\sethlcolor{usercolorname}
\usepackage{svg}
\usepackage{natbib}

\usepackage{multicol,multirow}
\newenvironment{mytabular}[1][1]{%
  \renewcommand*{\arraystretch}{#1}%
  \tabular%
}{%
  \endtabular
}

\begin{document}

\title[ChatGPT and U(X)]{ChatGPT and U(X): A Rapid Review on Measuring the User Experience}



\author*[1]{\fnm{Katie} \sur{Seaborn}}\email{seaborn.k.aa@m.titech.ac.jp}

\affil*[1]{\orgdiv{Department of Industrial Engineering and Economics}, \orgname{Institute of Science Tokyo}, \orgaddress{\street{2 Chome-12-1 Ookayama}, \city{Tokyo}, \postcode{152-8550}, \country{Japan}}}


\abstract{
ChatGPT, powered by a large language model (LLM), has revolutionized everyday human-computer interaction (HCI) since its 2022 release. 
While now used by millions around the world,  a coherent pathway for evaluating the user experience (UX) ChatGPT offers remains missing. In this rapid review ($N=58$), I explored how ChatGPT UX has been approached quantitatively so far. I focused on the independent variables (IVs) manipulated, the dependent variables (DVs) measured, and the methods used for measurement.
Findings reveal trends, gaps, and emerging consensus in UX assessments. 
This work offers a first step towards synthesizing existing approaches to measuring ChatGPT UX, urgent trajectories to advance standardization and breadth, and two preliminary frameworks aimed at guiding future research and tool development. 
I seek to elevate the field of ChatGPT UX by empowering researchers and practitioners in optimizing user interactions with ChatGPT and similar LLM-based systems.
}

\maketitle


\section{Introduction}

ChatGPT is a chatbot powered by a large language model (LLM) trained on the sum total of the known Internet. 
In over a two-year span---from its launch on November 30\textsuperscript{th}, 2022 to the time of reporting this review in December 2024---ChatGPT has generated worldwide interest~\citep{Ng_2024}, disrupted various creative and educational industries~\citep{tlili_what_2023,Amini_2025airimage}, gathered commercial and public investment~\citep{Ng_2024}, and sparked academic engagement from multiple angles~\citep{Shahriar2023}. Underlying ChatGPT is its Generative Pre-trained Transformer (GPT), an artificial intelligence (AI) language model, which enables a highly fluent and arguably persuasive~\citep{Jakesch2023} conversational experience, for novice and experienced users alike~\citep{gessinger_chatgpt_2025}. Researchers and practitioners alike have been awaiting this level of sophistication in agents of all kinds---chatbots, social robots, virtual assistants, and conversational user interfaces (CUI) in general~\citep{Clark2019}. The advent of ChatGPT voices in September 2023\footnote{\url{https://openai.com/index/chatgpt-can-now-see-hear-and-speak/}} hallmarked sound as a medium for LLM-based agents: from mere text to the natural human communication modality of speech~\citep{Seaborn2021voice}. As of August 2024, ChatGPT was used by over 200 million people each week\footnote{\url{https://www.reuters.com/technology/artificial-intelligence/openai-says-chatgpts-weekly-users-have-grown-200-million-2024-08-29/}}.

A plethora of work has turned to understanding how people perceive and interact with ChatGPT, hailing from an array of disciplines within and beyond the user-centred field of human-computer interaction (HCI)~\citep{Jakesch2023,skjuve_user_2023,kim_enhancing_2024}. This body of work, first characterized by preprints, is now starting to mature into peer-reviewed papers published to academic venues~\citep{Lo_2023}. Yet, like many novel and rapidly adopted technologies, ChatGPT needs a guiding hand to direct the course of work from a user experience (UX) angle~\citep{skjuve_user_2023,kim_enhancing_2024}. This is especially important given its purported fluency in comparison to now ``older'' models~\citep{desai_cuichi_2024,Lozi_2023fluent,Koco__2023,Lin_2024airtrust}. Moreover, ChatGPT can be deployed in the form provided by Open AI---the chatbot accessible via \url{https://chat.openai.com/}---or leveraged through the API\footnote{\url{https://platform.openai.com/docs/api-reference}} within other systems with relative ease~\citep{Mahmood2025gptva,Dong2023gptva}, effectively rendering its embodiment malleable. 
\citet{fabre2025makoto}, for example, employed ChatGPT as a storyteller agent with the voice of an older adult, unbeknownst to participants. 
At present, the what and how of UX with ChatGPT can be measured remains obscure. Moreover, new versions of the LLM and others like it are on the horizon, particularly ``fine-tuned'' models for specific purposes~\citep{Koco__2023,Zhou_2024tuning}. The time is ripe for a rapid review of the literature to steer the course for the next wave of human-centred design and research on ChatGPT and its kin, present and future.

To this end, I conducted a rapid review of the published work on measuring the UX with ChatGPT as a prototypical LLM-based chatbot. My research question (\textbf{RQ}) was: \emph{\textbf{How has UX been measured during user interactions with ChatGPT?}} 
I specifically asked: \textbf{RQ1}: \emph{What independent variables (IVs) have been manipulated?} \textbf{RQ2}: \emph{What dependent variables (DVs) have been measured?} and \textbf{RQ3}: \emph{What measures and forms of measurements have been used?} 
My goal was to reveal the current state of affairs, notably gaps and opportunities to set the stage for the next phase of work on ChatGPT UX. The contributions are: (i) A summary of approaches to measuring UX with ChatGPT; (ii) insights into emerging consensus, disagreements, gaps, and opportunities; and (iii) two frameworks for categorizing IVS and DVs to guide future studies. My aim was to assist researchers and others invested in ChatGPT UX with designing quantitative studies and potentially new, tailored instruments.


\section{Methods}

\begin{figure}[ht]
  \centering
  \includegraphics[width=\linewidth]{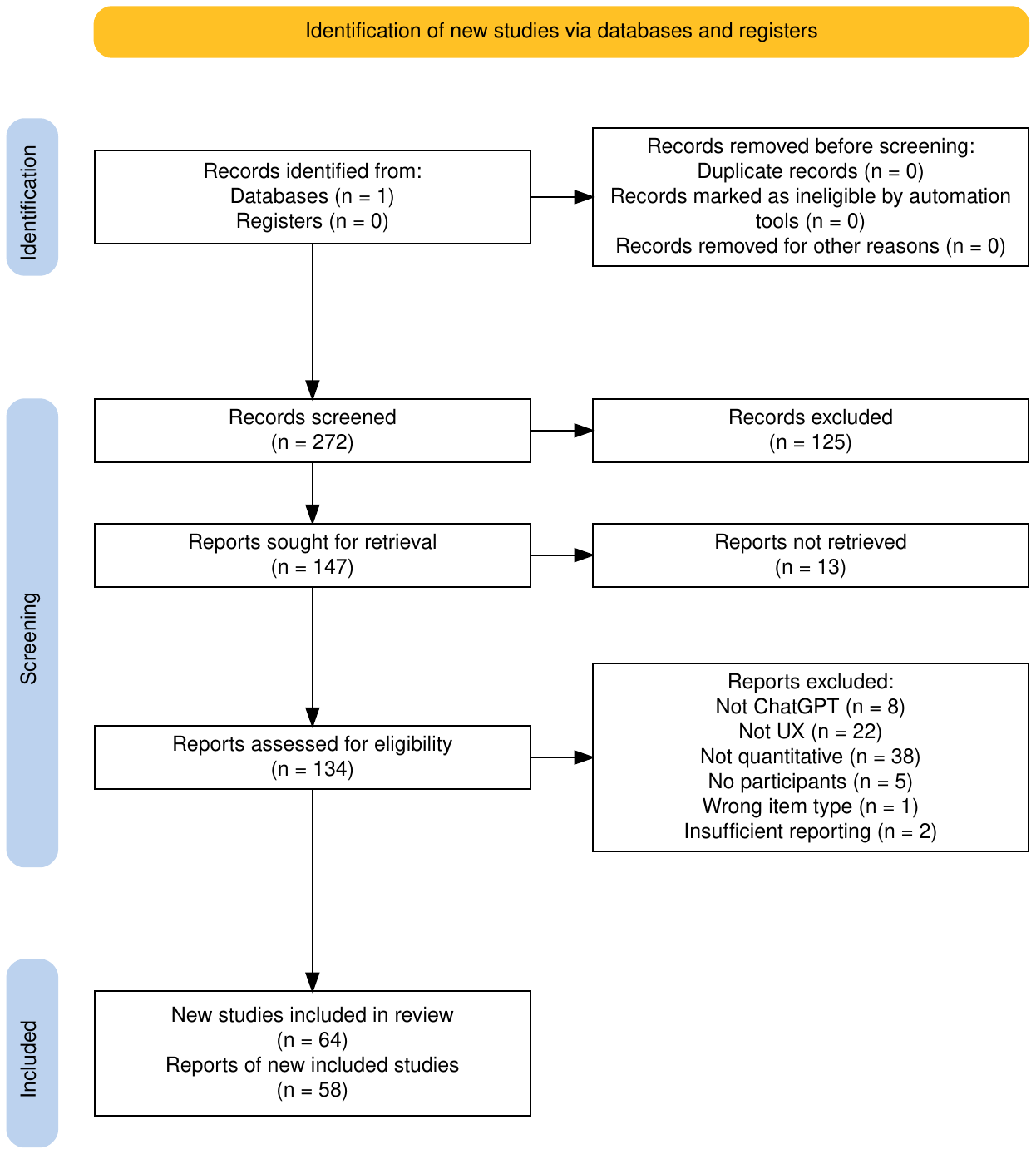}
  \caption{PRISMA flow diagram. Generated with the tool by \citet{Haddaway_2022rpris} (\url{https://estech.shinyapps.io/prisma_flowdiagram}).}
  \label{fig:a_flow}
\end{figure}

For the rapid review, I followed the world standard Cochrane  protocol~\citep{Garritty_2021rr} and the interim guidelines proposed by \citet{Stevens_2024prismarr}, aiming for the rigour recently promoted in HCI for literature reviews~\citep{Rogers_2023system}. Rapid reviews are a systematic way of capturing the current state of an emerging and dynamic literature. The rise and timeliness of ChatGPT and similar user-facing LLMs present it as an ideal subject. Rapid reviews generate a descriptive and narrative summary on a specific topic related to the dynamic subject; here, I focus on how UX has been approached as a measurable phenomenon of study. Typically, Cochrane-style reviews employ a standard protocol for reporting, notably the Preferred Reporting Items for Systematic Reviews and Meta-Analyses (PRISMA)~\citep{Page_2021prisma}. At present, however, the extension for rapid reviews is under development~\citep{Stevens_2024prismarr}. As suggested by the interim report~\citep{Stevens_2024prismarr}, I have used the PRISMA 2020 reporting guidelines, flow diagram (\autoref{fig:a_flow} in \ref{sec:a_flow}), and checklist (refer to \ref{sec:a_checklist}) as a baseline and added the decided items for the upcoming PRISMA-RR (PRISMA-Rapid Review) extension. These include, from Table 1, ``a priori-defined iterative methods,'' ```distinguishing the RR from a SR (scoping review),'' ``knowledge user involvement,'' ``authorship and corresponding author,'' ``acknowledgements,'' and ``peer review''~\cite[p. 3]{Stevens_2024prismarr}. I also adapted the protocol to suit the HCI literature, which, for instance, does not feature structured abstracts.

To address the second item relevant to rapid reviews above, this rapid review differs from an scoping review in several ways. The aim of the work was not to capture the broad scope of the field, but quickly gather the most pertinent examples that have been vetted by the community of practice. As described in \ref{sec:eligibility}, I excluded grey literature and unpublished work and only included work in the languages I know, limiting the scope in two ways. The work was also carried out by a single author, which is sufficient, if not common, for rapid reviews~\citep{Khangura_2012rr} but does not meet scoping review criteria~\citep{Munn_2018sr}. 

The rapid review was conducted in December 2024. The protocol was created a priori by the author, a researcher in HCI with published work on ChatGPT and expertise in review methodology, with no iterations (given solo authorship).
The protocol was preregistered on December 22\textsuperscript{nd}, 2024 at OSF
\footnote{\url{https://osf.io/wk5f6}}
 before the main search.

\subsection{Defining and Operationalizing UX}

The concept of UX is notoriously difficult to define and operationalize~\citep{Lallemand_2015ux,Hassenzahl_2006uxagenda,Law_2009ux}. \citet{Norman_1995ux} introduced the term as a complement to \textbf{usability}, with its narrow focus on user performance and safety. Soon after, \citet{Alben_1996exp} offered several criteria for designing ideal user interactions with systems for a ``successful and satisfying'' experience (p. 14). Indeed, UX has come to cover a ``broad range of fuzzy and dynamic concepts, including emotional, affective, experiential, hedonic, and aesthetic variables''~\cite[719]{Law_2009ux} within a ``fragmented'' theoretical landscape ''complicated by diverse theoretical models with different foci''~\cite[719]{Law_2009ux}. {\tt ISO 9241-210:2018} defines UX as a ``user's perceptions and responses that result from the use and/or anticipated use of a system, product or service''~\cite[§ 3.2.3]{ISO9241ux}. The standard goes on to describe the user response as ``emotions, beliefs, preferences, perceptions, comfort, behaviours, and accomplishments''~\cite[§ 3.2.3]{ISO9241ux} at all stages of engagement. UX arises from the combination of the user, the system, and the context~\citep{Lallemand_2015ux} and is relational, driven by internal and external factors to the user~\cite[§ 3.2.3]{ISO9241ux}.  Two large-scale surveys of professionals determined that UX is an individual experience that can be affected by the social context~\citep{Lallemand_2015ux, Law_2009ux}. Approaches to evaluating UX have been quantitative, qualitative, and mixed, with shifts over time and between industry and the academy~\citep{Lallemand_2015ux, Law_2009ux}. Recently, quantitative measurement involving self-report instruments on surveys, observation and performance metrics, and physiological responses has been centred~\citep{Inan_Nur_2021ux}, a trend noted in the related domains of voice UX~\citep{Clark2019,Seaborn_2021measur}. This guided my focus on quantitative operationalizations of ChatGPT UX as a first step.

The relationship between UX and usability deserves consideration. {\tt ISO 9241-210:2018} defines usability as ``extent to which a system, product or service can be used by specified users to achieve specified goals with effectiveness, efficiency and satisfaction in a specified context of use''~\cite[§ 2.13]{ISO9241usability}. While effectiveness (ability) and efficiency (speed) can be easily distinguished from UX, \textbf{satisfaction} is more tricky. {\tt ISO 9241-210:2018} defines satisfaction as the ``extent to which the user's physical, cognitive and emotional responses that result from the use of a system, product or service meet the user's needs and expectations,'' which also ``includes the extent to which the user experience that results from actual use meets the user's needs and expectations''~\cite[§ 3.1.14]{ISO9241usability}. 
Taking a cue from \citet{Lewis_2021usabilityux}, I include work that quantified \emph{subjective} experience and also treat broad measures of ``satisfaction'' as a distinct factor. I also recognize that, given the complex and diverse opinions on what constitutes UX~\citep{Lallemand_2015ux,Law_2009ux}, some factors may be categorized by others as distinct from UX, e.g., part of usability as a separate factor. In such cases, I follow the author/s.


\subsection{Eligibility Criteria (Inclusion and Exclusion Criteria)}
\label{sec:eligibility}

I included full user studies focusing on interactions with ChatGPT (as per the RQ). Likewise, each had to include at least one quantitative measure. For baseline quality, I excluded pilot studies, proposals and protocols, preprints, literature reviews,  grey literature (such as informal studies reported in news articles or social media), and papers lacking sufficient methodological detail. Only English- and Japanese-language papers were included (the languages that I knew).

\subsection{Information Sources, Search Queries, and Study Selection}

I searched Web of Science (per institutional access), which includes ACM Digital Library, IEEE Xplore, and other databases. I conducted the search on December 22\textsuperscript{nd}, 2024, using the following query structure:  {\tt (ALL=("user experience*" OR UX OR "customer experience*" OR "participant experience*" OR "user journey" OR "user interaction*" OR "user engagement*" OR "user perception*" OR "consumer perception*" OR "participant perception*" OR "experience design" OR "interaction flow" OR "immersive experience*" OR "user sentiment*" OR "customer sentiment*" OR "human factor*") AND ALL=(chatgpt OR "chat gpt"))}. 
I used the * qualifier to account for variations of each keyword, i.e., plurality, verb conjugations. I also filtered by {\tt Document Types (Article or Proceeding Paper or Early Access) and Language (English or Japanese)}. I alone screened the title, abstract, and keywords based on the eligibility criteria, then the full-text, conducting extractions alone. 

\subsection{Data Collection and Items}

I selected the items for extraction based on the RQs. These included: study descriptions (type of study, research design, context, participant demographics, GPT version); IVs; DVs, including measures, measurement, whether subjective/objective, response format, validation, if any, and origin, if any.

\subsection{Data Analysis}
I extracted metadata and then calculated descriptive statistics, such as counts and percentages per category. I then conducted two reflexive thematic analyses (RTA) to develop frameworks for the IV and DV data, following \citet{Braun_2006}. \citet{Clarke_2016} do not necessarily advocate for generalizabilty, arguing that the researcher's subjective position is informative knowledge. Likewise, RTA does not necessarily require multiple raters. My choice of RTA was guided by my solo authorship, the preliminary nature of this work, and my experience as a researcher in this area. 
I was influenced by similar work, notably \citet{Seaborn_2021measur} and \citet{Inan_Nur_2021ux}.
My frameworks are a form meta-research developed through grounded theory~\citep{Walker_2006gt,Glaser_2017gt}. I am epistemologically pluralistic and argue that these frameworks can be explored through more positivist or post-positivist approaches (Glaser style) and/or expanded in a social constructivist sense through community discourse (Strauss style)~\citep{Glaser_2017gt}.


\section{Results}

A total of \hl{272} reports were screened, leading to \hl{147} included for extraction and, after checking the full-text at this stage, ultimately \hl{58} reports covering \hl{64} studies were included for analysis (\autoref{fig:a_flow}). The full results are available in the open data set at \url{https://bit.ly/chatgptuxrr}. 

\subsection{Nature of the Studies}

\subsubsection{Demographics}
A total of 15,759 participants, comprising 5,348 men (33.9\%),	6,186 women (33.3\%),	17 of another gender (0.1\%), and 	17 who preferred not to report (0.1\%), were involved across all studies. The mean age was 30.5	($SD=8.0$). Most recruited from the general population (37, 58\%); there were also university students (13, 20\%), experts/professionals (9, 14\%), consumers (2, 3\%), grade school students (2, 3\%), and a child/caregiver combo (one case).

\subsubsection{ChatGPT Context}
Most studied GPT-3.5	(21,	36\%) and GPT-4 (13,	22\%); two used GPT-3. Notably, 21	(36\%) did not report the ChatGPT version. Three used the paid version of GPT-4, but most (55,	95\%) did not report on paid/free status. Two explored voice mode, while five explicitly did not and 51 (88\%) did not report either way.
Most used the chatbot UI provided by Open AI (38,	66\%), while 18 (31\%) created their own chatbots, robots (2, 3\%), or virtual characters (one case). 
ChatGPT was explored in general use cases (26,	45\%), education	(13,	22\%), marketing/business	(8,	14\%), medical	(6, 10\%), research	(2,	3\%), entertainment (storytelling, games)	(1,	2\%); one each were on driving and assembly line work.

\subsubsection{Study Context}
Most were online (33,	52\%), with some in lab (22,	34\%) and in field (9,	14\%). Most were experiments (28,	44\%), surveys	(19,	30\%), and user studies	(12,	19\%), with some focus groups (3, 5\%) and two sentiment analyses.

\subsection{What independent variables (IVs) have been manipulated? (RQ1)}

28 (44\%) of the 64 studies had IVs, with
34 IVs across these studies. The IV framework is provided in \autoref{tab:ivs}.

\begin{table}[]
\caption{IV framework; counts and percentages are relative to the 46 analytical codes across the 28 studies that employed IVs.}
\label{tab:ivs}
\begin{tabular}{p{50pt}p{55pt}p{95pt}p{125pt}l}
\toprule
\textbf{Category} & \textbf{Subcategories} & \textbf{Examples} & \textbf{Sources} & \textbf{Scope} \\
\midrule
\multirow{2}*{\shortstack[l]{User \\Characteristics}} & User Group & Region & \citep{fujii_silver-tongued_2024} & 1 (2\%) \\
 & Experience Level & Novices vs. experienced users & \citep{fabio_ai-based_2024,bertaglia_closing_2023,gessinger_chatgpt_2025} & 3 (7\%) \\
\midrule
Presentation & Form Factor & Fantasy narrative game vs. non-gamified pedagogical virtual agent; robot vs. ChatGPT & \citep{steenstra_engaging_2024,jarsve_exploring_2024,zhang_human_2023} & 3 (7\%) \\
 & Identity & Persona based on pronoun use; agent personality; brand identity; human and AI configurations & \citep{bertaglia_closing_2023,fujii_silver-tongued_2024,park_traditional_2024,steenstra_engaging_2024,zhang_human_2023} & 5 (11\%) \\
 & Framing & Disclosure of ChatGPTs limitations; presence of misinformation and when; serious game or educational context; disclosure of informational capacity & \citep{amaro_believe_2024,leiser_hill_2024,khurana_why_2024,zhang_human_2023}  & 6 (13\%) \\
 & Agent Comparison & ChatGPT vs. other agent using ChatGPT or not; ChatGPT vs. humans & \citep{de_lima_ai-powered_2025,amaro_believe_2024,ha_clochat_2024,masson_directgpt_2024,leiser_hill_2024,khurana_why_2024,jarsve_exploring_2024,zhang_human_2023} & 10 (22\%) \\
\midrule
Interaction Style & Communicative Style & Timing of suggestions; authoritativeness; listening support; hesitation and self-editing; explanation style & \citep{kuang_enhancing_2024,metzger_empowering_2024,zhao_exploring_2024,pasternak_3d_2024,zhou_beyond_2024,huang_chatbot_2024,amaro_believe_2024,pafa_unraveling_2024} & 10 (22\%) \\
 & Fluency & Answer correctness; dictation ability; conversational complexity or sophistication & \citep{de_lima_ai-powered_2025,huang_chatbot_2024,lin_rambler_2024,pafa_unraveling_2024} & 4 (9\%) \\
 & Degree of Aid & Programming aid level; presence of aid & \citep{sun_would_2024,khurana_why_2024} & 2 (4\%) \\
\midrule
\multirow{2}*{\shortstack[l]{Interaction\\ Modality}} & Query Form & Fixed verbal responses vs. natural responses & \citep{ye_improved_2023} & 1 (2\%) \\
 & Paradigm & Direct interaction vs. text-based typing & \citep{masson_directgpt_2024} & 1 (2\%) \\
 \bottomrule
\end{tabular}
\end{table}

\subsection{What dependent variables (DVs) have been measured? (RQ2)}

143 DVs were reported across the 64 studies.  The DV framework is provided in \autoref{tab:dvs}. DVs were sometimes divided into factors or constructs, or not given names (in which case I relied on the item/s). Some DVs also crossed several categories. Hence, the counts may be greater than the number of sources.

\begin{table}[]
\caption{DV framework; counts and percentages are relative to the 232 analytical codes across the 143 DVs.}
\label{tab:dvs}
\begin{tabular}{p{50pt}p{55pt}p{55pt}p{145pt}l}
\toprule
\textbf{Category} & \textbf{Subcategories} & \textbf{Examples/Sources} &  & \textbf{Scope} \\
\midrule

Appeal & Engagement & \multicolumn{2}{p{200pt}}{Stimulation; interactivity; captured attention; use over time; usage frequency; interest} &  20 (9\%)\\
& & \multicolumn{3}{p{250pt}}{\citep{lieb_student_2024,lent_chat_2024,de_lima_ai-powered_2025,kim_unveiling_2024,pasternak_3d_2024,sun_would_2024,fabio_ai-based_2024,zhou_beyond_2024,yan_does_2024,kim_decisions_2023,jo_subscription_2024,ha_clochat_2024,isaza-giraldo_prompt-gaming_2024,zhao_language_2024,yu_can_2024,jo_understanding_2023,choudhury_exploring_2024,zhang_human_2023}} \\

 & Intents & \multicolumn{2}{p{200pt}}{Intent to use; willingness to use; intent to continue use} &  16 (7\%) \\
 & & \multicolumn{3}{p{250pt}}{\citep{choudhury_investigating_2023,shahsavar_user_2023,chen_conversational_2024,pasternak_3d_2024,tossell_student_2024,sun_would_2024,fabio_ai-based_2024,quan_young_2024,kim_decisions_2023,jo_subscription_2024,park_traditional_2024,abadie_shared_2024,jo_understanding_2023,choudhury_exploring_2024,zhang_human_2023}} \\
 
 & Values & \multicolumn{2}{p{200pt}}{Beneficial; threat to society; societal level positive attitudes; potential for bias; novel ideas} & 8 (3\%) \\
 & & \multicolumn{3}{p{250pt}}{\citep{tossell_student_2024,fabio_ai-based_2024,ouaazki_generative_2024,quan_young_2024,yan_does_2024,gessinger_chatgpt_2025,rodriguez_leveraging_2024}} \\
\midrule

\multirow{2}*{\shortstack[l]{Attitudes\\ \& Sentiments}} & Trust & \multicolumn{2}{p{200pt}}{Credibility; dependability; security; privacy; transparent; forthcoming about hallucinations}  & 40 (17\%) \\
& & \multicolumn{3}{p{250pt}}{\citep{choudhury_investigating_2023,xing_factors_2024,lent_chat_2024,kuang_enhancing_2024,metzger_empowering_2024,zhang_exploring_2024,de_lima_ai-powered_2025,kim_unveiling_2024,ye_improved_2023,tossell_student_2024,fabio_ai-based_2024,salah_chatting_2024,bertaglia_closing_2023,ouaazki_leveraging_2023,yan_does_2024,gessinger_chatgpt_2025,amaro_believe_2024,choudhury_impact_2024,pafa_unraveling_2024,leiser_hill_2024,khurana_why_2024,rodriguez_leveraging_2024,han_determinants_2024,jo_understanding_2023,choudhury_exploring_2024}} \\

 & Acceptance & \multicolumn{2}{p{200pt}}{Novelty; risk-reward appraisal; desirability; agreement with recommendations} & 21 (9\%) \\
 & & \multicolumn{3}{p{250pt}}{\citep{lent_chat_2024,romero-rodriguez_use_2023,metzger_empowering_2024,de_lima_ai-powered_2025,shahsavar_user_2023,chen_conversational_2024,sun_would_2024,salah_chatting_2024,bertaglia_closing_2023,fujii_silver-tongued_2024,yan_does_2024,huang_chatbot_2024,ha_clochat_2024,choudhury_impact_2024,zhao_language_2024,han_determinants_2024,choudhury_exploring_2024}} \\
 
 & Affect & \multicolumn{2}{p{200pt}}{Apprehension; enjoyment; positive and negative emotions; comfort} & 29 (13\%)\\
 & & \multicolumn{3}{p{250pt}}{\citep{han_determinants_2024,zhao_language_2024,pafa_unraveling_2024,isaza-giraldo_prompt-gaming_2024,xu_public_2024,ha_clochat_2024,kimmel_enhancing_2024,huang_chatbot_2024,gessinger_chatgpt_2025,yan_does_2024,ngo_does_2024,ouaazki_generative_2024,bertaglia_closing_2023,fabio_ai-based_2024,tossell_student_2024,pasternak_3d_2024,chen_conversational_2024,kim_unveiling_2024,de_lima_ai-powered_2025,zhao_exploring_2024,metzger_empowering_2024,lent_chat_2024,lieb_student_2024}}  \\
\bottomrule
\end{tabular}
\end{table}

\begin{table}[]
\caption{DV framework, continued; counts and percentages are relative to the 232 analytical codes across the 143 DVs.}
\label{tab:dvs_2}
\begin{tabular}{p{50pt}p{55pt}p{55pt}p{145pt}l}
\toprule
\textbf{Category} & \textbf{Subcategories} & \textbf{Examples/Sources} &  & \textbf{Scope} \\
\midrule

\multirow{2}*{\shortstack[l]{Agentic \\Qualities}} & Anthromorphism & \multicolumn{2}{p{200pt}}{Growth; human-likeness; natural; genderedness; agedness; self-efficacy} & 14 (6\%) \\
& & \multicolumn{3}{p{250pt}}{\citep{xing_factors_2024,metzger_empowering_2024,chen_conversational_2024,fujii_silver-tongued_2024,zhou_beyond_2024,gessinger_chatgpt_2025,kim_decisions_2023,jo_subscription_2024,ha_clochat_2024,jarsve_exploring_2024}} \\

 & Personality & \multicolumn{2}{p{200pt}}{Friendly; benevolent; openness; loyal; warm or cold; personality type} & 12 (5\%)\\
 & & \multicolumn{3}{p{250pt}}{\citep{park_traditional_2024,ha_clochat_2024,huang_chatbot_2024,gessinger_chatgpt_2025,zhou_beyond_2024,holderried_generative_2024,ouaazki_leveraging_2023,fabio_ai-based_2024,tossell_student_2024,pasternak_3d_2024,chen_conversational_2024}} \\
 
 & Intelligence & \multicolumn{2}{p{200pt}}{Logical inference; creativity; decision-making; cognitive complexity; reasoning style; flexible} & 35 (15\%) \\
 & & \multicolumn{3}{p{250pt}}{\citep{xing_factors_2024,lieb_student_2024,kim_unveiling_2024,shahsavar_user_2023,chen_conversational_2024,pasternak_3d_2024,tossell_student_2024,fabio_ai-based_2024,miah_user_2024,bertaglia_closing_2023,ouaazki_leveraging_2023,quan_young_2024,zhou_beyond_2024,yan_does_2024,gessinger_chatgpt_2025,amaro_believe_2024,jo_subscription_2024,ha_clochat_2024,yu_can_2024,nielsen_usefulness_2024,han_determinants_2024,jo_understanding_2023,choudhury_exploring_2024}} \\
\midrule

Relationality & Social Presence & \multicolumn{2}{p{200pt}}{(Social) interactivity; casual conversation; cooperative; friendly; social factors} & 12 (5\%)\\
& & \multicolumn{3}{p{250pt}}{\citep{jo_understanding_2023,ha_clochat_2024,kim_decisions_2023,huang_chatbot_2024,gessinger_chatgpt_2025,fabio_ai-based_2024,pasternak_3d_2024,kim_unveiling_2024,de_lima_ai-powered_2025,lent_chat_2024,xing_factors_2024}}  \\

 & Helpfulness & \multicolumn{2}{p{200pt}}{Assists; improved ability; taught new skills; relevance} & 25 (11\%) \\
 & & \multicolumn{3}{p{250pt}}{\citep{lieb_student_2024,bertaglia_closing_2023,ouaazki_leveraging_2023,ouaazki_generative_2024,quan_young_2024,yan_does_2024,huang_chatbot_2024,jo_subscription_2024,kimmel_enhancing_2024,ha_clochat_2024,choudhury_impact_2024,pafa_unraveling_2024,yu_can_2024,rodriguez_leveraging_2024,nielsen_usefulness_2024,han_determinants_2024,jo_understanding_2023,choudhury_exploring_2024}} \\
 
 & Personalization & \multicolumn{2}{p{200pt}}{Response personalization; persona customization; understands ``me''; personalized attention; stereotyping; individual impact} & 18 (8\%) \\
 & & \multicolumn{3}{p{250pt}}{\citep{jo_understanding_2023,park_traditional_2024,leiser_hill_2024,yu_can_2024,ha_clochat_2024,jo_subscription_2024,amaro_believe_2024,gessinger_chatgpt_2025,ouaazki_leveraging_2023,salah_chatting_2024,fabio_ai-based_2024,pasternak_3d_2024,kim_unveiling_2024,zhao_exploring_2024,abdelkader_chatgpts_2023}} \\
\bottomrule
\end{tabular}
\end{table}

\begin{table}[]
\caption{DV framework, continued; counts and percentages are relative to the 232 analytical codes across the 143 DVs.}
\label{tab:dvs_3}
\begin{tabular}{p{50pt}p{55pt}p{55pt}p{145pt}l}
\toprule
\textbf{Category} & \textbf{Subcategories} & \textbf{Examples/Sources} &  & \textbf{Scope} \\
\midrule

Usability &  & \multicolumn{2}{p{200pt}}{Ease of use; accuracy; convenience; information quality; system quality; triability; understandability; difficulty; effective; user-friendly; improved work efficiency; error handling; mental demand} & 81 (35\%) \\
& & \multicolumn{3}{p{250pt}}{\citep{xing_factors_2024,abdelkader_chatgpts_2023,lieb_student_2024,lent_chat_2024,kuang_enhancing_2024,zhao_exploring_2024,de_lima_ai-powered_2025,kim_unveiling_2024,shahsavar_user_2023,chen_conversational_2024,pasternak_3d_2024,sun_would_2024,fabio_ai-based_2024,miah_user_2024,bertaglia_closing_2023,ouaazki_generative_2024,ouaazki_leveraging_2023,alonso_seamless_2024,quan_young_2024,holderried_generative_2024,zhou_beyond_2024,yan_does_2024,gessinger_chatgpt_2025,huang_chatbot_2024,amaro_believe_2024,kim_decisions_2023,jo_subscription_2024,kim_enhancing_2024,ha_clochat_2024,masson_directgpt_2024,isaza-giraldo_prompt-gaming_2024,lin_rambler_2024,pafa_unraveling_2024,zhao_language_2024,yu_can_2024,leiser_hill_2024,khurana_why_2024,steenstra_engaging_2024,rodriguez_leveraging_2024,nielsen_usefulness_2024,han_determinants_2024,jo_understanding_2023,choudhury_exploring_2024}} \\
\midrule

Satisfaction &  & \multicolumn{2}{p{200pt}}{Overall satisfaction; timing preference satisfaction; satisfied/pleased/contented; nice/likable; likelihood of recommending; UX as satisfaction; time and number of attempts as satisfaction} & 20 (9\%) \\
& & \multicolumn{3}{p{250pt}}{\citep{zhang_human_2023,choudhury_exploring_2024,jo_understanding_2023,rodriguez_leveraging_2024,steenstra_engaging_2024,pafa_unraveling_2024,choudhury_impact_2024,ha_clochat_2024,jo_subscription_2024,kim_decisions_2023,amaro_believe_2024,huang_chatbot_2024,zhou_beyond_2024,miah_user_2024,barambones_chatgpt_2024,chen_conversational_2024,kim_unveiling_2024,zhao_exploring_2024,kuang_enhancing_2024,abdelkader_chatgpts_2023}} \\
\bottomrule
\end{tabular}
\end{table}

\subsection{What measures and forms of measurements have been used? (RQ3)}

Most DVs used  self-reports	(131,	92\%), with some using metrics	(8,	6\%) and observation	(4,	3\%); there were no physiological measures. The most common response formats were 7-point Likert scales	(55,	38\%), 
5-point Likert scales	(48,	34\%),
counts/numeric data	(11, 8\%), and
4-point Likert scales	(10	7\%). 
Most self-report measures used more than 1 item		(180,	78\%; $M=4, SD=3.9$), while 50 relied on one-item measures (22\%). Other options included checklists, yes/no questions, semantic differential scales, and other Likert scales. Three did not report on the response format.

The majority of measures were novel instruments (70,	49\%), with some existing	(51,	36\%), some modified	(20,	14\%), and two unreported. 37	(26\%) were validated, while 59	(41\%) were not, and 	47	(33\%) were unreported. Similarly,	56	(39\%) checked reliability, while 48 (34\%) did not and 39	(27\%) did not report. Across the 58 reports, 17 (29\%) created a new instrument, but of these, nine (53\%) were not validated.

Existing, validated instruments included: User Experience Questionnaire (UEQ)~\citep{Laugwitz_2008}; Short Version of the User Experience Questionnaire (UEQ-S)~\citep{Schrepp_2017}; Standardized User Experience Percentile Rank Questionnaire (SUPR-Q)~\citep{sauro2015supr}; System Usability Scale (SUS)~\citep{brooke1996sus}; Chatbot Usability Scale~\citep{Borsci_2021}; the Unified Theory of Acceptance and Use of Technology (UTAUT)~\citep{Venkatesh_2003}; Unified Theory of Acceptance and Use of Technology 2 (UTUAT2)~\citep{Venkatesh_2012}; Chatbot Usability Questionnaire (CUQ)~\citep{Holmes_2019}; Partner Modelling Questionnaire (PMQ)~\citep{doyle2023partner}; negative attitudes towards robots scale (NARS)~\citep{Nomura_2006}; artificial-social-agent questionnaire (ASA)~\citep{Fitrianie_2022}; General Attitudes towards Artificial Intelligence Scale (GAAIS)~\citep{Schepman_2022}; General Attitude Towards Robots Scale (GAToRS)~\citep{Koverola_2022}; human-computer trust scale~\citep{Gulati_2019}; AI trust score~\citep{Wang_2021trust}; Multi-Dimensional Measure of Trust (MDMT), Version 2~\citep{Ullman_2019}; Critical Reflective Assessment (CRA)~\citep{Anghel2021cra}; Behaviors from Intergroup Affect and Stereotypes-Treatment Scale (BIAS-TS)~\citep{Sibley_2011}; and mental effort scale~\citep{Paas_1992}.

Novel, validated instruments designed for ChatGPT included: UX in AI chat systems~\citep{xing_factors_2024}; influence on customer experience in digital marketing~\citep{abdelkader_chatgpts_2023}; attitudes on policies towards ChatGPT~\citep{yan_does_2024}; subscription intentions for ChatGPT plus~\citep{jo_subscription_2024}; trust~\citep{choudhury_impact_2024}; priori acceptance~\citep{abadie_shared_2024}; approach behavior~\citep{han_determinants_2024}; and AI tool engagement~\citep{jo_understanding_2023}.




\section{Discussion}

ChatGPT has taken the world by storm, enabling new forms of UX and setting a new standard for chatbots and other systems that speak with us. Measuring the UX ChatGPT offers reflects the nascency of this technology and its general uptake by regular people (\textbf{overarching RQ}). Few IVs have been explored relative to, for instance, UX in general~\citep{Inan_Nur_2021ux} and the related domain of voice UX~\citep{Seaborn_2021measur}. Communication style and comparison to other agents are prevalent. Some validated instruments have been applied or created to explore ChatGPT UX across different contexts and in pursuit of certain UX-related DVs, but these are mixed with a plethora of unvalidated or single-item measures. Standardization of instruments and consolidation of DVs will be needed before we can fully grasp ChatGPT UX through meta-analyses and systematic reviews. The reader is encouraged to peruse the open data set for inspiration: \url{https://bit.ly/chatgptuxrr}


A diversity of variables are being manipulated to explore the UX people have with ChatGPT (\textbf{RQ1}). \emph{User characters} are a crucial element. That varying levels of expertise has been studied---from novices with no experience to those who have become experts with daily use~\citep{bertaglia_closing_2023,fabio_ai-based_2024,gessinger_chatgpt_2025}---is not unexpected. Work is starting to dig deeper, considering sociolingusitic elements relevant to fluent LLM engagements, such as the comparison of Japanese regions in \cite{fujii_silver-tongued_2024}. This work demonstrates the importance of not only basic fluency in a given language, but catering to specific regions: accents and expectations in the social elements linked to language~\citep{carmichael2016place}.

ChatGPT is also an interactive technology: one part UI, one part agent. \emph{Presentation} is crucial. ChatGPT can be used in its native chatbot state or employed in game worlds~\citep{steenstra_engaging_2024} or social robots~\citep{jarsve_exploring_2024} and represent different identities~\citep{park_traditional_2024} and personas~\citep{fujii_silver-tongued_2024}, even simply by self-identifying or using certain pronouns. \emph{Framing} is linked to trust~\citep{Lin_2024airtrust}, especially given ongoing issues with ChatGPT's propensity to hallucinate. \cite{zhang_human_2023} provides a template for running controlled experimental designs to test the degree and effect of disclosing use of ChatGPT. Indeed, comparisons to other agents~\citep{amaro_believe_2024,ha_clochat_2024,leiser_hill_2024,khurana_why_2024} and humans~\citep{amaro_believe_2024,zhang_human_2023} may be ideal, not only for trust but other UX factors.

\emph{Interaction style} and \emph{modality} are related variables with distinct manipulations. ChatGPT, as a communicative partner, has pushed the community to go deeper with measures of communicative style, fluency, and degree of aid. \cite{zhou_beyond_2024}, for instance, show how sublime factors like hesitation and self-reflection potentially modulate impressions of nonverbal fluency, key to social agent design~\citep{urakami_nonverbal_2023}. The basic fluency of ChatGPT as an LLM has also opened up approaches to query forms, as demonstrated by \cite{ye_improved_2023}, and modality, as shown by \cite{masson_directgpt_2024} through a direct, drag-and-drop interaction with ChatGPT that blends conversation and imagery. This work, revealed through the present rapid review, reflects a combination of traditional HCI and UX manipulations alongside emerging possibilities enabled by the power of the LLM models and advances in UI technology. However, there was a notable gap: \emph{voice}. ChatGPT voice mode, at the time of the research, was a premium feature, perhaps explaining the dearth of work. \cite{huang_chatbot_2024} explored ChatGPT voice in a driving assistant, a clear use case when the user's eyes should be on the road. Likewise, \cite{yu_can_2024} used voice mode during language learning, where listening and speaking is key. Text and voice modalities offer different UX and may be used together~\citep{Clark2019,Seaborn2021voice}. The efficacy, fluency, and use cases for ChatGPT voices are sure to form the next wave of research on ChatGPT UX.



Dependent variables (\textbf{RQ2})---what UX phenomena were measured---and how these were measured (\textbf{RQ3}) were diverse in scope and novelty. The degree and nature of ChatGPT's \emph{appeal} was explored from multiple fronts. ChatGPT can stimulate and capture attention in the moment, but more significantly, over time and across contexts, from the classroom~\citep{yu_can_2024,steenstra_engaging_2024} to the workplace~\citep{choudhury_exploring_2024,jo_understanding_2023}. A growing body of work has begun to explore intent and willingness to use~\citep{abadie_shared_2024,jo_subscription_2024}, generating subjective frameworks largely based on expectations that should be tested in longitudinal studies. \emph{Dis}engagement based on mismatched values and threat potential to individuals and societies has also been explored. \cite{yan_does_2024} offers a compelling evaluation of attitudes towards ChatGPT's impact on jobs, now and in the future. More work is needed on these topics to fully assess whether ChatGPT should be deployed and with whom by industry and context of use.

A broader range of \emph{attitudes and sentiments} have been studied, notably trust, acceptance, and affect. The rapid review on this body of work reveals a steadfast grounding in established theories, like the Unified Theory of Acceptance and Use of Technology (UTAUT)~\citep{Venkatesh_2003}, as well as the development of new descriptive frameworks and instruments tailored to ChatGPT and similar technologies, such as the AI Trust Score~\citep{Wang_2021trust}. The spread of measures (notably single-item and unvalidated instruments) and varying scope (from individual factors, like perceptions of transparency and level of personalization) indicate that ChatGPT UX is variegated and tied to context-specific concerns. This was true for affect, as well. For instance, \cite{isaza-giraldo_prompt-gaming_2024} narrowed the focus on feelings of fun, given the game context, while \cite{kimmel_enhancing_2024} focused on feelings of frustration, given the objective of understanding the UX of error messages. What is likely to emerge is an array of tailored general UX instruments alongside specific measures geared around specific user groups, contexts of use, and UX objectives.

ChatGPT generally takes the form of an agent with \emph{agentic qualities} and a social \emph{relationality} to the user. Subsequently, degree of anthropomorphism or humanlikeness emerged as a grand category. The more specific concepts of personality and intelligence were also highly relevant within the corpus. While many used unvalidated and single-item measures, some drew from extant and recent materials. \cite{gessinger_chatgpt_2025} provides a case study in their use of the Partner Model Questionnaire (PMQ)~\citep{doyle2023partner}, a new and validated instrument of agent communicative ability. This instrument uses semantic differential scales---such as Human-like/machine-like and Personal/generic---to evaluate a rich array of qualities related to anthropomorphism and contextualized to communication scenarios. Such humanlike qualities also link to the social experience: social presence, helpfulness, and personalization. These qualities are multifaceted and may be measured together. \cite{pasternak_3d_2024}, for instance, deployed the artificial-social-agent or ASA scale~\citep{Fitrianie_2022}, which asked users to evaluate the relevance of statements like ``The [agent] understands me.'' While we may think of intelligence and sociality as distinct, the the reviewed body of work suggests that \emph{social intelligence} will be a key dimension in assessments of ChatGPT UX and likely other sophisticated, user-facing AI~\citep{Fan_2022socialint}.

\emph{Usability}, to a large extent, and \emph{satisfaction}, to a lesser extent, were also explored across the reviewed papers. UX is directly relevant to each of these factors~\citep{ISO9241ux,ISO9241usability,Norman_1995ux,Hassenzahl_2006uxagenda,Lewis_2021usabilityux}. Usability measures tended to be more observational and objective, . But some relied on subjective self-reports of usefulness~\citep{ouaazki_generative_2024,ouaazki_leveraging_2023}, ease of use~\citep{kim_unveiling_2024,ouaazki_generative_2024,amaro_believe_2024,zhao_language_2024,steenstra_engaging_2024}, efficiency~\citep{kim_unveiling_2024}, and accuracy~\citep{miah_user_2024,holderried_generative_2024}. Some, like \cite{leiser_hill_2024, alonso_seamless_2024,masson_directgpt_2024} used the simple but established subjective self-report-based System Usability Scale or SUS~\citep{Bangor_2008sus}. Satisfaction measures, as per the construct, were more often subjective. Exceptions existed, such as \cite{miah_user_2024} operationalizing number of attempts to use ChatGPT as a measure of satisfaction. Altogether, this demonstrates an emerging landscape of established measures productively used to understand ChatGPT UX, with customizations and even new measure and instruments developed to address its particularities.


An important point of discussion is the state of affairs on the quality of the instruments used (\textbf{RQ3}). In parallel with the findings for voice UX~\citep{Seaborn_2021measur}, the results on how ChatGPT UX has been measured indicate urgent action items for improving the quality, assessing the validity, and evaluating the reliability of the measures and forms of measurement being employed. Uptake of validated, multi-item instruments like the UEQ~\citep{Laugwitz_2008} was concurrent with two other trends: in this cohort, only one-quarter of studies deployed validated instruments and nearly one-quarter were one-item measures. \cite{Allen_2022oneitem} discuss the issue with nuance. While some research has shown that single-item measures are as reliable and valid as multi-item instruments~\citep{Ang_2017}, a ``measure'' of doubt is needed given such issues as publication biases against null results~\citep{Allen_2022oneitem}. Assessing the reliability of one-item measures also requires more time and effort, such as by using test-retest reliability to ensure that the phenomenon linked to the item can be stably measured over time and across participants~\citep{Allen_2022oneitem,Polit_2014}. Ensuring validity also poses difficulties for solo item measures. Convergent validity---by comparison to an un/validated multi-item instrument---predictive validity---by way of theory-driven hypotheses---and concurrent validity---combining the previous two by showing that the solo item is correlated with theory and equivalent to a multi-items measure---requires more involved procedures and likely multiple trials. Finally, a single item may not capture the full complexity of the factor under study~\citep{Allen_2022oneitem,Bergkvist_2007}. On this last point, the vast array of DVs discovered in this rapid review alongside the wealth of measures---subjective and objective; validated and unvalidated; tested for reliability or not; one-item or multifaceted; established or novel---sketch out a complex picture for UX with ChatGPT. This state of affairs can be interpreted as a launching point for the next phase in measuring ChatGPT UX with rigour and intricacy.

\subsection{Limitations and Future Work}
I restricted the scope to ChatGPT. Several other LLM-based chatbots and chatbot-style systems exist, including some that may leverage the same or similar models used for ChatGPT (like Perplexity). Future systematic reviews should include such systems. Additionally, I did not aim to report on findings from the studies, largely because the work is still nascent and generalizable results may be premature. As the field matures, systematic reviews and meta-syntheses should focus on the nature of ChatGPT UX and impacts on various user groups. Finally, future systematic reviews should involve multiple screeners, extractors, and raters for quality~\citep{Rogers_2023system}.

\section{Conclusion: A Research Agenda}

ChatGPT represents a shift in chatbot-based HCI experiences: not only in agent fluency, but also in terms of impact on our daily lives. We are entering a world where such LLMs are joining us at various levels in society 
as collaborators and companions. 
I offer these crucial next steps for this field of study and practice:

\begin{itemize}
    \item \textbf{Standardized instruments}: ChatGPT is a landmark technology, having pushed the quality of chatbot-based UX forward greatly. A variety of user groups are taking up ChatGPT; various contexts of use are emerging; and models geared towards certain purposes and people are also on the rise. Some have made commendable efforts towards understanding ChatGPT UX at larger scales and in pursuit of instrument and/or theory development. The 74\% validity gap in the work so far needs to be redressed with instrument uptake and validation.

    \item \textbf{Social intelligence measures}: ChatGPT represents a new standard for chatbots, but also other user-facing AI. The reviewed body of work hints at an emerging factor within the purview of anthropomorphism: a link between agent sociality and intelligence. As yet, no work has explored social intelligence theoretically or formally in measures or instruments. This offers a gap that may be filled by existing or novel instruments. Still, each has been explored in isolation, as captured by the \emph{relationality} theme (23\%) and \emph{intelligence} sub-theme (15\%). The review by \cite{Fan_2022socialint} may guide future work. As they argue, social intelligence should not be considered in isolation, but is contextual and linked to other factors, notably social processing, like shared perceptions of social events and social inference~\citep{shu2020adventures} and mentalizing, or inferring values and beliefs~\citep{Gonz_lez_2021}, paralinguistics, such as politeness~\citep{Ribino_2023airpolite}, and nonverbal communication~\citep{urakami_nonverbal_2023}, perhaps most relevant to ChatGPT embodiments that have a visible form to indicate gaze and gestures.

    \item \textbf{Fine-tuned evaluation by GPT version}: Over one-third cast a wide net in terms of GPT version and model. This is a critical issue given the differences in quality, setup, and experience. \citet{barambones_chatgpt_2024}, for instance, found that 3.5 and 4 operated differently, notably requiring special prompts to ensure the same conversational flow between GPT versions. Also, virtually no one, with the exception of \citet{jo_subscription_2024}, reported on whether the paid version was used, which may be critical given speed and base model differences. OpenAI reportedly had over 11 million paying users in September 2024\footnote{\url{https://www.theinformation.com/articles/openai-coo-says-chatgpt-passed-11-million-paying-subscribers}}, suggesting the potential of a substantial but hidden special user group. O1\footnote{\url{https://openai.com/o1/}} may be especially key for HCI and broader fields in computing, given its capacity to act as a code-generator and programming assistant. We will need standardized ways of measuring the experience each model offers to fully grasp the minutiae of the UX and whether these fine-grained differences matter.

    \item \textbf{Voice mode}: Only two studies explicitly assessed voice mode, despite its release over a year ago (September 2023\footnote{\url{https://openai.com/index/chatgpt-can-now-see-hear-and-speak/}}). Systematic reviews on voice UX have revealed how speech is a powerful medium that can influence user perceptions, attitudes, and behaviour~\citep{Seaborn2021voice,Seaborn_2021measur,Clark2019}. Yet, I was unable to find a single study on ChatGPT voice UX. Most did not make a clear demarcation between the text-based version and use of voice mode, if any existed, with the exception of \citet{kim_unveiling_2024}. This will be a critical path of research going forward, especially for those who are blind or low-vision (B/LV)~\citep{Kuzdeuov_2024blv} and anyone else who is better served by non-visual modalities. 
\end{itemize}

\newpage
\section*{Statements and Declarations}

\textbf{Acknowledgements and Funding:} None declared.\\

\noindent \textbf{Use of AI:} I acknowledge editorial use of ChatGPT.\\

\noindent \textbf{Competing Interests:} None declared.\\

\noindent \textbf{Data Availability:} The data set is available here: \url{https://bit.ly/chatgptuxrr}

\newpage
\begin{appendices}

\section{Appendix}

\subsection{PRISMA Checklist}
\label{sec:a_checklist}


\emph{Original version of the PRISMA 2020, with revisions by the author (Katie Seaborn) based on the interim PRISMA-RR guidance by \citet{Stevens_2024prismarr}, from:} Page M.J., McKenzie J.E., Bossuyt P.M., Boutron I., Hoffmann T.C., Mulrow C.D., et al. The PRISMA 2020 statement: An updated guideline for reporting systematic reviews. BMJ 2021;372:n71. doi:
\href{https://doi.org/10.1136/bmj.n71}{10.1136/bmj.n71}

\FloatBarrier

\begin{table}[!ht]
\caption{PRISMA 2020 Checklist.}
\label{table:prisma}
{\small
\begin{mytabular}[1.2]{|>{\raggedright}p{2.6cm}|>{\centering}p{0.75cm}|>{\raggedright}p{7.2cm}|p{3cm}|}

\hline
\rowcolor{aquablue}
Section and Topic & Item  & Checklist Item & Reported on Page \# \\
\hline 
\multicolumn{4}{l}{TITLE} \\
\hline
Title & 1 & Identify the report as a systematic review. & \hl{p. 1}\\
\hline \hline
\multicolumn{4}{l}{ABSTRACT} \\
\hline
Abstract & 2 &See the PRISMA 2020 for Abstracts Checklist. & Refer below\\
\hline \hline
\multicolumn{4}{l}{INTRODUCTION} \\
\hline
Rationale & 3 & Describe the rationale for the review in the context of existing knowledge. &  \hl{p. 2} \\ 
\hline
Objectives & 4 & Provide an explicit statement of the objective(s) or question(s) the review addresses. & \hl{p. 2} \\
\hline \hline
\multicolumn{4}{l}{METHODS} \\
\hline
Eligibility Criteria & 5  & Specify the inclusion and exclusion criteria for the review and how studies were grouped for the syntheses & \hl{p. 4} \\ 
\hline
  Information sources & 6 & Specify all databases, registers, websites, organisations, reference lists and other sources searched or consulted to identify studies. Specify the date when each source was last searched or consulted. & \hl{p. 5} \\
 \hline
Search strategy & 7  & Present the full search strategies for all databases, registers and websites, including any filters and limits used. & \hl{p. 5} \\ 
\hline
Selection process &  8&  Specify the methods used to decide whether a study met the inclusion criteria of the review, including how many reviewers screened each record and each report retrieved, whether they worked independently, and if applicable, details of automation tools used in the process. & \hl{p. 5} \\

\hline \hline
\end{mytabular}}
\end{table}

\begin{table}[!ht]
\caption{PRISMA 2020 Checklist, continued.}
\label{table:prisma2}
{\small
\begin{mytabular}[1.2]{|>{\raggedright}p{2.6cm}|>{\centering}p{0.75cm}|>{\raggedright}p{7.2cm}|p{3cm}|}

\hline
\rowcolor{aquablue}
Section and Topic & Item  & Checklist Item & Reported on Page \# \\
\hline
\multicolumn{4}{l}{METHODS, Cont.} \\

\hline
Data collection process & 9 &  Specify the methods used to collect data from reports, including how many reviewers collected data from each report, whether they worked independently, any processes for obtaining or confirming data from study investigators, and if applicable, details of automation tools used in the process. & \hl{p. 5} \\ 
\hline
 \multirow{2}{2.5cm}{Data items}& 10a  &  List and define all outcomes for which data were sought. Specify whether all results that were compatible with each outcome domain in each study were sought (e.g. for all measures, time points, analyses), and if not, the methods used to decide which results to collect.& \hl{p. 5} \\
\cline{2-4}
 & 10b & List and define all other variables for which data were sought (e.g. participant and intervention characteristics, funding sources). Describe any assumptions made about any missing or unclear information. & \hl{p. 5} \\ 
 
\hline
Study risk of bias assessment &11  & Specify the methods used to assess risk of bias in the included studies, including details of the tool(s) used, how many reviewers assessed each study and whether they worked independently, and if applicable, details of automation tools used in the process. & N/A; not for meta-analysis \\
 \hline
Effect measures  &  12 & Specify for each outcome the effect measure(s) (e.g. risk ratio, mean difference) used in the synthesis or presentation of results &  N/A; not for meta-analysis \\ 
\hline
\multirow{3}{2.5cm}{Synthesis methods} & 13a & Describe the processes used to decide which studies were eligible for each synthesis (e.g. tabulating the study intervention characteristics and comparing against the planned groups for each synthesis (item \#5)). & \hl{p. 5} \\
 \cline{2-4}
 & 13b & Describe any methods required to prepare the data for presentation or synthesis, such as handling of missing summary statistics, or data conversions. & \hl{N/A; not needed} \\ 
\cline{2-4}
 & 13c & Describe any methods used to tabulate or visually display results of individual studies and syntheses. & \hl{p. 5} \\

\hline \hline
\end{mytabular}}
\end{table}

\begin{table}[!ht]
\caption{PRISMA 2020 Checklist, continued.}
\label{table:prisma3}
{\small
\begin{mytabular}[1.2]{|>{\raggedright}p{2.6cm}|>{\centering}p{0.75cm}|>{\raggedright}p{7.2cm}|p{3cm}|}

\hline
\rowcolor{aquablue}
Section and Topic & Item  & Checklist Item & Reported on Page \# \\
\hline
\multicolumn{4}{l}{METHODS, Cont.} \\

\hline

\multirow{3}{2.5cm}{Synthesis methods} & 13d & Describe any methods used to synthesize results and provide a rationale for the choice(s). If meta-analysis was performed, describe the model(s), method(s) to identify the presence and extent of statistical heterogeneity, and software package(s) used. & \hl{p. 5} \\ 
\cline{2-4}
 & 13e & Describe any methods used to explore possible causes of heterogeneity among study results (e.g. subgroup analysis, meta-regression). & N/A; not for meta-analysis \\
\cline{2-4}
 & 13f & Describe any sensitivity analyses conducted to assess robustness of the synthesized results. & N/A; not for meta-analysis \\ 
\hline
Reporting bias assessment & 14 & Describe any methods used to assess risk of bias due to missing results in a synthesis (arising from reporting biases). & N/A; not for meta-analysis \\
 \hline
Certainty assessment & 15 & Describe any methods used to assess certainty (or confidence) in the body of evidence for an outcome. & N/A; not for meta-analysis \\

\hline
A priori-defined iterative methods & RR-1 &  Report whether an iterative process (ideally specified in the protocol) was used, such as decision-making on methodology or inclusion during the conduct of the review to meet the timeline. & \hl{p. 5} \\
\hline
Distinguishing the RR from an SR & RR-2 & Indicate what aspects of the conduct or process that would differ from an SR. & \hl{p. 3} \\
\hline
Knowledge user involvement & RR-3 & Describe what knowledge users (eg, policymakers, patients, guideline developers, clinicians) were involved in the development of the RR, specifying the stage(s) and the nature of involvement. & \hl{N/A; not topical} \\
\hline \hline

\multicolumn{4}{l}{RESULTS} \\
\hline
 \multirow{2}{2.5cm}{Study selection}& 16a & Describe the results of the search and selection process, from the number of records identified in the search to the number of studies included in the review, ideally using a flow diagram. & \hl{p. 5;} \ref{sec:a_flow} \\
 \cline{2-4}
 & 16b & Cite studies that might appear to meet the inclusion criteria, but which were excluded, and explain why they were excluded. & N/A \\
 \hline 
 Study Characteristics & 17 &  Cite each included study and present its characteristics. & Refer to tables \\
\hline
 Risk of bias in studies & 18  & Present assessments of risk of bias for each included study. & N/A; not for meta-analysis \\
\hline \hline
\end{mytabular}}
\end{table}

\begin{table}[!ht]
\caption{PRISMA 2020 Checklist, continued.}
\label{table:prisma4}
{\small
\begin{mytabular}[1.2]{|>{\raggedright}p{2.6cm}|>{\centering}p{0.75cm}|>{\raggedright}p{7.2cm}|p{3cm}|}

\hline
\rowcolor{aquablue}
Section and Topic & Item  & Checklist Item & Reported on Page \# \\
\hline
\multicolumn{4}{l}{RESULTS, Cont.} \\
\hline
 Results of individual studies & 19 &  For all outcomes, present, for each study: (a) summary statistics for each group (where appropriate) and (b) an effect estimate and its precision (e.g. confidence/credible interval), ideally using structured tables or plots.& N/A; not for meta-analysis \\ 
\hline 
 \multirow{4}{2.5cm}{Results of syntheses} & 20a & For each synthesis, briefly summarise the characteristics and risk of bias among contributing studies & \hl{pp. 5--11} \\
 \cline{2-4}
& 20b & Present results of all statistical syntheses conducted. If meta-analysis was done, present for each the summary estimate and its precision (e.g. confidence/credible interval) and measures of statistical heterogeneity. If comparing groups, describe the direction of the effect. & \hl{pp. 5--11} \\
\cline{2-4}
& 20c & Present results of all investigations of possible causes of heterogeneity among study results. & N/A; not for meta-analysis \\
\cline{2-4}
& 20d & Present results of all sensitivity analyses conducted to assess the robustness of the synthesized results. & N/A; not for meta-analysis \\
\hline
 Reporting biases & 21 & Present assessments of risk of bias due to missing results (arising from reporting biases) for each synthesis assessed. & N/A; not for meta-analysis \\
 \hline
 Certainty of evidence &22 & Present assessments of certainty (or confidence) in the body of evidence for each outcome assessed. &N/A; not for meta-analysis \\
 
 \hline \hline
\multicolumn{4}{l}{DISCUSSION} \\
\hline
\multirow{4}{2.5cm}{Discussion}&23a  & Provide a general interpretation of the results in the context of other evidence. & \hl{pp. 11--14} \\
\cline{2-4}
 & 23b & Discuss any limitations of the evidence included in the review. & \hl{p. 14} \\
 \cline{2-4}
 & 23c & Discuss any limitations of the review processes used. & \hl{p. 14} \\
 \cline{2-4}
 &  23d & Discuss implications of the results for practice, policy, and future research. & \hl{p. 14} \\

\hline \hline
\end{mytabular}}
\end{table}

\begin{table}[!ht]
\caption{PRISMA 2020 Checklist, continued.}
\label{table:prisma5}
{\small
\begin{mytabular}[1.2]{|>{\raggedright}p{2.6cm}|>{\centering}p{0.75cm}|>{\raggedright}p{7.2cm}|p{3cm}|}

\hline
\rowcolor{aquablue}
Section and Topic & Item  & Checklist Item & Reported on Page \# \\
\hline
\multicolumn{4}{l}{OTHER INFORMATION} \\
\hline
\multirow{3}{2.5cm}{Registration and protocol} & 24a  &  Provide registration information for the review, including register name and registration number, or state that the review was not registered.& \hl{p. 3} \\
\cline{2-4}
 &  24b&Indicate where the review protocol can be accessed, or state that a protocol was not prepared. & \hl{p. 3} \\
 \cline{2-4}
 &  24c& Describe and explain any amendments to information provided at registration or in the protocol. & \hl{p. 3} \\
 \hline
 Support & 25 & Describe sources of financial or non-financial support for the review, and the role of the funders or sponsors in the review. & \\ \hline
 Competing interests &  26& Declare any competing interests of review authors.& \hl{Acknowledgements} \\ \hline
  Availability of data, code and other materials & 27 & Report which of the following are publicly available and where they can be found: template data collection forms; data extracted from included studies; data used for all analyses; analytic code; any other materials used in the review. & \hl{p. 3, p. 11, p. 16} \\ 
\hline
Authorship and corresponding author & RR-4 & List those who contributed sufficiently to meet authorship requirements. Provide contact information for the corresponding author or organisational representative. & \hl{p. 1} \\

\hline
Acknowledgements & RR-5 & List those who contributed to the development and conduct the work but do not meet authorship requirements. & \hl{N/A} \\
\hline
Peer review & RR-6 & Indicate whether peer review was undertaken during the preparation of the report and by whom (eg, methodologist or content expert and whether internal or external to producing organisation). & \hl{p. 3} \\
\hline \hline
\end{mytabular}}
\end{table}

\begin{table}[!ht]
\caption{PRISMA 2020 Abstract Checklist.}
\label{table:prisma6}
{\small
\begin{mytabular}[1.2]{|>{\raggedright}p{2.6cm}|>{\centering}p{0.75cm}|>{\raggedright}p{7.2cm}|p{3cm}|}

\hline
\rowcolor{aquablue}
Section and Topic & Item  & Checklist Item & Reported on Page \# \\
\hline 
\multicolumn{4}{l}{TITLE} \\
\hline
Title & 1 & Identify the report as a systematic review. & p. 1 (as rapid review) \\
\hline \hline
\multicolumn{4}{l}{BACKGROUND} \\
\hline 
Objectives & 2 &Provide an explicit statement of the main objective(s) or question(s) the review addresses. & \hl{p. 1} \\
\hline \hline 
\multicolumn{4}{l}{METHODS} \\
\hline 
Eligibility criteria &3 &Specify the inclusion and exclusion criteria for the review. & \hl{N/A; disciplinary standards} \\
\hline 
Information sources & 4&Specify the information sources (e.g. databases, registers) used to identify studies and the date when each was last searched. & \hl{N/A; disciplinary standards} \\
\hline 
Risk of bias& 5& Specify the methods used to assess risk of bias in the included studies. & \hl{N/A; disciplinary standards} \\
\hline 
Synthesis of results & 6& Specify the methods used to present and synthesise results.& \hl{p. 1} \\
\hline \hline 
\multicolumn{4}{l}{RESULTS} \\
\hline 
Included studies & 7& Give the total number of included studies and participants and summarise relevant characteristics of studies.& \hl{p. 1} \\
\hline 
Synthesis of results & 8&Present results for main outcomes, preferably indicating the number of included studies and participants for each. If meta-analysis was done, report the summary estimate and confidence/credible interval. If comparing groups, indicate the direction of the effect (i.e. which group is favoured). & \hl{p. 1} \\
\hline \hline 
\multicolumn{4}{l}{DISCUSSION} \\
\hline 
Limitations of evidence& 9 &Provide a brief summary of the limitations of the evidence included in the review (e.g. study risk of bias, inconsistency and imprecision). &\hl{N/A; disciplinary standards} \\
\hline 
Interpretation&  10& Provide a general interpretation of the results and important implications.&\hl{p. 1} \\
\hline \hline 
\multicolumn{4}{l}{OTHER} \\
\hline 
Funding& 11& Specify the primary source of funding for the review.&\hl{N/A; disciplinary standards} \\
\hline 
Registration& 12& Provide the register name and registration number.& \hl{N/A; disciplinary standards}\\
\hline \hline
\end{mytabular}}
\end{table}

\FloatBarrier

\newpage
\subsection{PRISMA Flow Diagram}
\label{sec:a_flow}

\end{appendices}

\newpage

\bibliography{refs,refs-zotero}

\end{document}